\documentclass[osajnl,twocolumn,showpacs,superscriptaddress,10pt]{revtex4-1} %% use 11pt for Applied Optics
\usepackage{amsmath,amssymb,graphicx}
\begin{document}
\title{Observing angular deviations in light beam reflection via weak measurements.}
\author{G. Jayaswal}
\author{G. Mistura}
\author{M. Merano}\email{Corresponding author: michele.merano@unipd.it}
\affiliation{Dipartimento di Fisica e Astronomia G. Galilei, Universit\`{a} degli studi di Padova, via Marzolo 8, 35131 Padova, Italy}

\begin{abstract}An optical analog of the quantum weak measurement scheme proved to be very useful for the observation of optical beam shifts. Here we adapt the weak value amplification method for the observation of the angular Goos-H$\rm\ddot{a}$nchen shift. We observe this effect in the case of external air-dielectric reflection, the more fundamental case in which it occurs. We show that weak measurements allow for a faithful amplification of the effect at any angle of incidence, even at the Brewster angle of incidence.
\end{abstract}

\ocis{270.0270, 050.1940, 270.0270.}

\maketitle %% required
A beam of light, reflected at a planar interface, does not follow perfectly the ray optics prediction. A small angular deviation of the Law of Reflection has been predicted for a physical light beam when this is regarded as the implementation of a ray \cite{Bertoni73, Boerner74, Chan85}. Experimental proofs of this phenomenon were reported recently \cite{Merano09}. This angular deviation of the beam axis that occurs only in the case of partial reflection has been named angular Goos-H$\rm\ddot{a}$nchen (AGH) shift, in analogy with the Goos-H$\rm\ddot{a}$nchen (GH) shift \cite{Goos47}, which is a positional shift of the beam centre relative to its geometrical optics position. The AGH and the GH occur in the plane of incidence. Out-of-plane shifts such as the Imbert-Fedorov (IF) shift \cite{Imbert72, Fedorov55} and the Spin Hall effect of Light (SHEL) \cite{Onoda04, Bliokh06, Kwiat08} have been observed as well. Conservation laws imply that these phenomena affect also the transmitted beam and not only the reflected one \cite{Bliokh06}. 

Weak measurements proved to be very convenient for the observation of some of these phenomena. Hosten and Kwiat used this experimental approach to measure the exact magnitude of the SHEL \cite{Kwiat08}. Their observation closed a theoretical debate on this subject. This experiment was done for a light beam transmitted at an air-glass interface. The same experimental approach was used to observe the SHEL in optical reflection \cite{Qin09} as well as in a plasmonic system \cite{Gorodetski12}. The observations of the GH \cite{Jayaswal13} and the IF shift \cite{Jayaswal14} via a weak measurement scheme were reported as well. This exploitation of a well known quantum weak measurement technique for the observation of beam shifts required a detailed theoretical treatment \cite{Aiello08}. The result of this analysis was a complete classical description of both the observed phenomena and the experimental scheme used to observe them \cite{Toppel13, Dennis12, Gotte12, Gotte13}.

Here we report the weak measurement of the AGH.  We observe the AGH of a Gaussian light beam at an air-glass interface. We show that weak measurements allow for a faithful amplification of this shift for any angle of incidence, and in particular even at the Brewster angle ($\theta_B$) where a $p$ polarized beam experiences a dispersive resonance for the AGH \cite{Merano09}. 

Technically speaking we measure the weak value of the polarization of a light beam. The AGH shift acts as the weak measuring effect, and an analyzer post-selects the final polarization state. The experimental results of this procedure can be interpreted as an amplification scheme for the AGH \cite{Aharonov88, Aharonov90}. As it will be clear later, the choice of the pre-selected state is subtle here because it is affected by the amplitude reflection coefficients.

We adopt the quantum mechanical description developed in \cite{Toppel13}. We use the Jones calculus to describe the polarization of the light in combination with the bra-ket notation where  $\left\langle p\left|\right.\right.$= (1, 0) is the \textit{p} polarization and $\left\langle s\left|\right.\right.$= (0, 1) is the \textit{s} polarization. The quantum operator describing the AGH effect is given by
\renewcommand{\arraystretch}{1} 
\begin{equation}
AGH=
\left[\begin{array}{cc}
\Theta_p & 0 \\
0 & \Theta_s \\	
\end{array}\right]
%\end{equation}
%\begin{equation}
\end{equation} 

%\renewcommand{\arraystretch}{2} 
%\begin{equation}
%AGH=
%\left[\begin{array}{cc}
%-i\dfrac{\partial\hspace{1 mm}ln\hspace{0.5 mm}r_p}{\partial\hspace{1 mm}\theta} & 0 \\
%0 & -i\dfrac{\partial\hspace{1 mm}ln\hspace{0.5 mm}r_s}{\partial\hspace{1 mm}\theta} \\	
%\end{array}\right]
%%\end{equation}
%%\begin{equation}
%\end{equation} 
where 
\begin{equation}
\Theta_p=-i\dfrac{\partial\hspace{1 mm}ln\hspace{0.5 mm}r_p}{\partial\hspace{1 mm}\theta}
\: \: \: \: \: \: \: \: \: \: \: \: \Theta_s= -i\dfrac{\partial\hspace{1 mm}ln\hspace{0.5 mm}r_s}{\partial\hspace{1 mm}\theta}
\end{equation}  
Here $r_p$ and $r_s$ are the amplitude reflection coefficients and $\theta$ is the angle of incidence. At $\theta_B$ the expression for $\Theta_p$ must be corrected. For angles of incidence close to $\theta_B$ ($\mid \theta-\theta_B \mid < 3\theta_0$, where $\theta_0$ is the beam angular aperture of the Gaussian beam) \cite{Merano09} we have:
\begin{equation}
\Theta_p=-i\dfrac{\theta-\theta_B}{\Lambda(\theta-\theta_B)^2+\alpha}
\end{equation}  
where:
\begin{equation}
\alpha=\dfrac{1}{2}+\dfrac{2n^2}{(1+n^2)^4}
\end{equation}
This expression comes from the finite divergence of a Gaussian $p$ polarized light beam that unavoidably implies a small admixture of $s$-polarization \cite{Merano09, AielloXiv09}. For a $p$ polarized Gaussian sheet beam $\alpha =1/2$ in this expression.  
For reasons that will be clear later, it is convinient to write the matrix $AGH$ in this form:
\begin{eqnarray}
AGH = \frac{1}{2}(\Theta_p+\Theta_s)\boldsymbol{I} +\frac{1}{2}(\Theta_p-\Theta_s)\boldsymbol{\sigma_3} 
\end{eqnarray}
where $\bold{I}$ is the identity matrix and $\bold{\sigma_3}$ is the Pauli matrix:
\begin{equation}
\boldsymbol{\sigma_3}=
\left[\begin{array}{cc}
1 & 0 \\
0 & -1\\	
\end{array}\right]
\end{equation}  
Since $AGH$ is diagonal in the $p$, $s$ basis we can not pre-select the polarization to be $p$ or $s$. When a ray of light is reflected at an interface its polarization is preserved only for $p$ or $s$ states. If it is not possible to pre-select $p$ or $s$ polarization then the reflection matrix (ref. \cite{Toppel13} formula (22a)):
\renewcommand{\arraystretch}{1} 
\begin{equation}
F=
\left[\begin{array}{cc}
r_p & 0 \\
0 & r_s \\	
\end{array}\right]
%\end{equation}
%\begin{equation}
\end{equation} 
must be considered also in our formalism. Form ref. \cite{Toppel13} (formulas (25, 57a and 57 b) it is clear that the reflection matrix is part of the pre-selection process for the input polarization. If the polarization of the light beam incident on the dielectric surface is $\left|\psi\right\rangle$ then the pre-selected polarization for the weak measurement scheme is $\left|\gamma\right\rangle=F\left|\psi\right\rangle$.

Our procedure for the observation of the AGH effect runs as follow. We pre-select $\left\langle\gamma\left|\right.\right.$=(1/$\sqrt{2}$, -1/$\sqrt{2}$). We then post-select two final polarization states: 
\begin{equation}
\left\langle\phi\right| = (\frac{1\pm\epsilon}{\sqrt{2}}, \frac{1\mp\epsilon}{\sqrt{2}})
\end{equation} 
(where $\epsilon$ is a small angle). The weak value of the AGH matrix is:
\begin{equation}
\frac{\left\langle\phi\left|AGH\right| \gamma \right\rangle }{\left\langle \phi | \gamma \right\rangle}
= \frac{1}{2}(\Theta_p+\Theta_s)\pm\frac{1}{2\epsilon}(\Theta_p-\Theta_s)
\end{equation}
where the addend divided by epsilon comes form $\bold{\sigma_3}$, while the other addend comes from  $\bold{I}$. We note that in the case of air-glass external reflection this weak value is purely imaginary because the reflection coefficients are reals. The centroids position $\left\langle x\right\rangle$  of the post-selected beams in the direction parallel to the plane of incidence is proportional to the this weak value according to the rule \cite{Aiello08}:
\begin{equation}
\left\langle x\right\rangle = -\frac{\lambda}{2\pi}\frac{z}{z_{0}}\bigg( (\frac{1}{2}(\Theta_p+\Theta_s)\pm\frac{1}{2\epsilon}(\Theta_p-\Theta_s)\bigg)
\end{equation}
where $z$ is the propagation distance, $z_0$ is the Rayleigh range and $\lambda$ is the wavelength of the light.

\begin{figure}
\includegraphics[scale = 0.3]{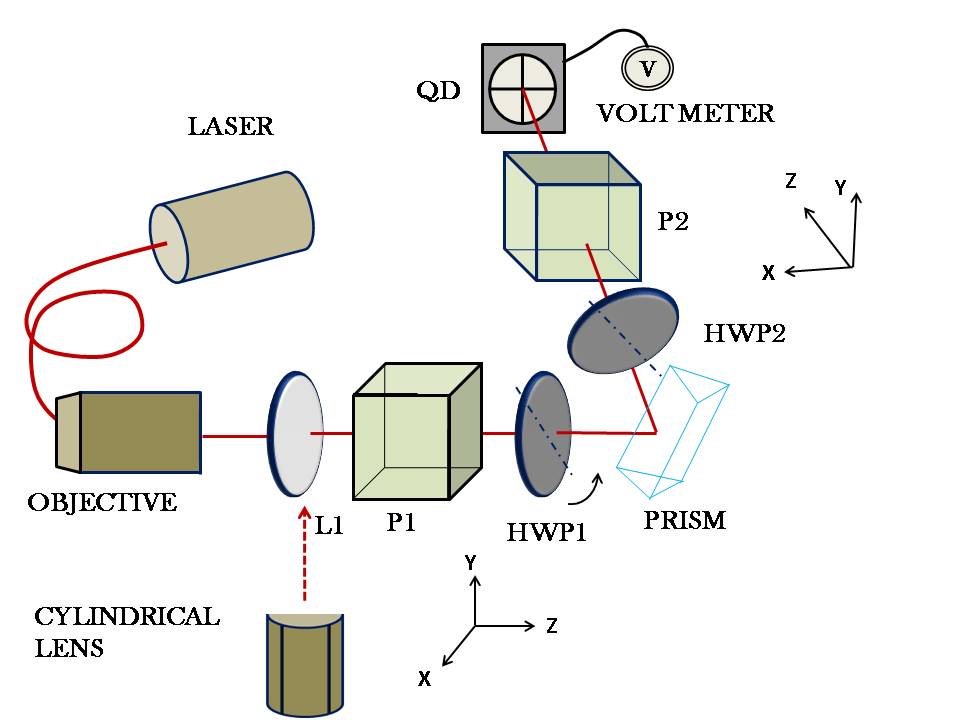}
\caption{Set up for the weak measurement of the angular Goos-H$\rm\ddot{a}$nchen shift. L: lens. P1, P2: Glan-Thompson polarizers. HWP1, HWP2: half wave plates. QD: Quadrant detector. The pre-selected state is fix by HWP1 and the reflection matrix F. The polarization analyser (HWP2 and P2) post-selects the final polarization state. The prism is mounted on a precision rotation stage (not shown).}
\end{figure}

We now turn to the experimental set-up shown in Fig. 1. We used a collimated Gaussian light beam, from a fiber-pigtailed laser diode, with $\lambda$ = 826 nm. A lens L1 (focal lengh=20 cm or 7.5 cm) transforms the waist parameter $w_0$ to a desired value at a z-position (chosen as the origin). The beam is polarized by means of a Glan-Thompson polarizing prism P1. A half wave plate (HWP1) is used to fix the polarization of the light beam incident on a $45\,^{\circ}$-$90\,^{\circ}$-$45\,^{\circ}$ BK7 prism and externally reflected from it. After this an analyser, composed of a half wave plate (HWP2) and a Glan-Thompson polarizing prism P2, is used to post-selects the desired polarization states. A quadrant detector (QD), placed at a distance z from the beam waist, and connected to a nanovoltmeter measures the beam displacements.
 
Measurements are performed in this way. P2 is set to projects on the state $\left\langle+45\,^{\circ}\right|$=(1/$\sqrt{2}$, 1/$\sqrt{2}$) and the optical axis of the HWP2 is initially set to be at $+45\,^{\circ}$ from the x axis:
\begin{equation}
HWP2=
\left[
\begin{array}{cc}
0 & 1\\
1 & 0 \\	
\end{array}\right].
\end{equation} 
Polarizer P1 simply linearly polarizes the light from the laser. HWP1 turns the linear polarization from P1 to be $\left\langle\psi\right|=(\frac{A}{\sqrt{2}r_p}, -\frac{A}{\sqrt{2}r_s}$) (where A is an irrelevant normalization factor that is simplified in (9)). Because the reflection coefficients in external reflection are real number with modulus smaller than one, it is always possible to set a state of polarization like this with the HWP1. After reflection the polarization of the beam is given by $\left|\gamma\right\rangle$. So, in practice, we set HWP1 by turning the orientation of its optical axis until we have the minimum power transmitted to the QD. We then turn HWP2 of a small angle $\pm\epsilon/2$:
\begin{equation}
HWP2=
\left[
\begin{array}{cc}
\mp\epsilon & 1\\
1 & \pm\epsilon \\	
\end{array}\right].
\end{equation} 
In this way we post-select the two final states:
\begin{equation}
\left\langle +45\,^{\circ} \right| HWP2 = \left\langle\phi\right|.
\end{equation}
The QD measures the relative distance $\Delta$ of the beam centroids for the two post-selected states. From formula (10) this is given by 
\begin{equation}
\Delta = -\frac{\lambda}{2\pi}\frac{z}{z_{0}}\bigg( \frac{\Theta_p-\Theta_s}{\epsilon}\bigg).
\end{equation}
The observed angular shift $\Theta$ is given by
\begin{equation}
\Theta = \frac{\Delta}{z}\cdot\frac{1}{\epsilon}
\end{equation}
where the first term is the expression for the angular shift of a $p$ polarized beam with respect to an $s$ polarized one and it is exactly equal to the one that can be obtained from formula (4) of ref. \cite{Aiello08} or, close to $\theta_B$, from formulas (4 and 5) of ref. \cite{Merano09}. The second term is the weak value amplification factor.
\begin{figure}
\includegraphics{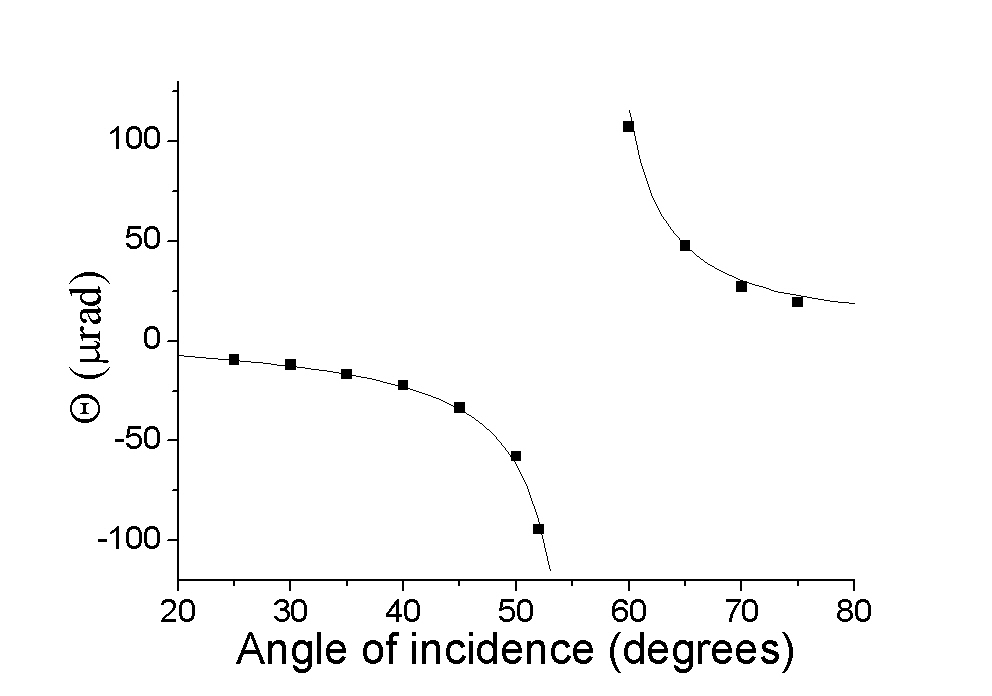}
\caption{Angular Goos-H$\rm\ddot{a}$nchen shift for a $p$ polarized Gaussian beam with respect to a $s$ polarized one. Solid dots are experimental data, the line is the theoretical prediction. The agreement in between theory and experiment is excellent as long as we are far from Brewster.}
\end{figure}  

Figure 2 shows our results for angles of incidence far from $\theta_B$. The line represents the theoretically predicted AGH of a $p$ polarized beam with respect to a $s$ polarized one. Solid dots are the measured experimental data divided by the amplification factor $1/\epsilon$. In the set of measurements reported $w_0$= 70$\rm \mu m$, $\epsilon$ = 0.097 rad and the distance z of the QD from the beam waist is 30 cm. The agreement in between theory and experiment is excellent. This is the first experimental result of this papers and it shows that, for $\theta$ far from $\theta_B$, weak measurements allow for a faithful amplification of the angular shift. 

\begin{figure}[h]
\includegraphics{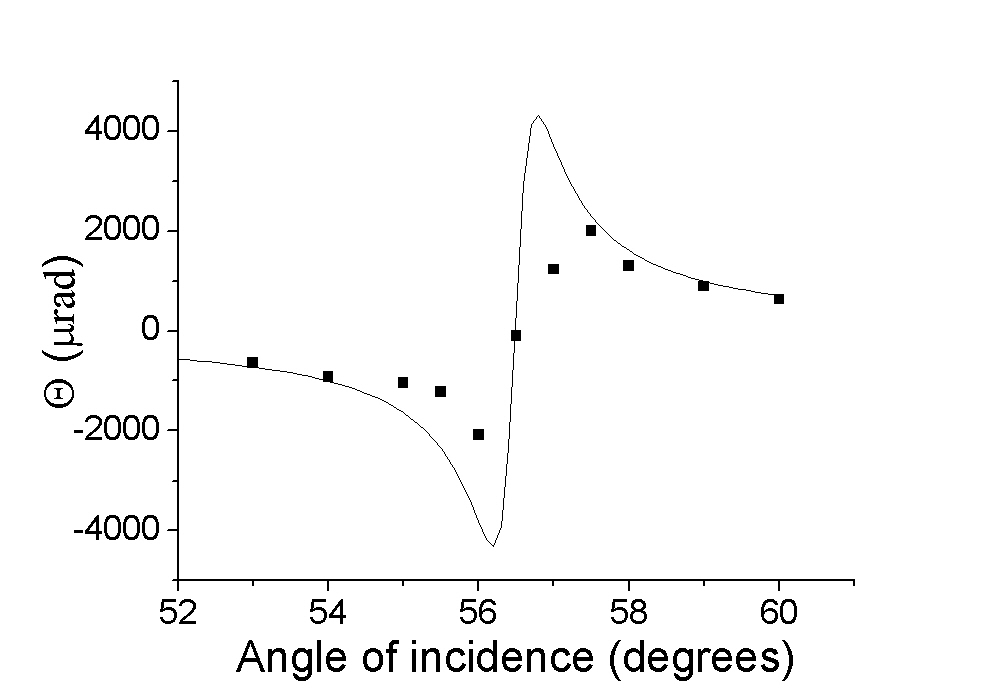}
\caption{Angular Goos-H$\rm\ddot{a}$nchen shift for a $p$ polarized Gaussian beam with respect to a $s$ polarized one. Solid dots are experimental data, the line is the theoretical prediction. The agreement in between theory and experiment is not anymore good at Brewster.}
\end{figure}  

For angles of incidence close to $\theta_B$ weak measurements are succesful as well (Fig. 4), if we use a sheet Gaussian beam instead of a Gaussian one. In this way we avoid the Brewster cross polarization effect that alters the reflection process and the weak measurement scheme as well. Figure 3 and Fig. 4 compare our experimental results for these two cases.

Figure 3. reports our experimemtal results for a Gaussian beam where the relevant parameters are $w_0$= 28$\rm \mu m$, $\epsilon$ varies in the range 0.06 - 0.26 rad (the biggest value was chosen at $\theta_B$ where the reflected intensity has a minimum) and z = 25 cm. We observe the expected dispersive resonance, but the experimental results have not the right magnitude. This is due to the fact that at $\theta_B$ a Gaussian $p$ polarized beam generates a two-mode beam with both a dominant and a cross polarized component (Brewster cross-polarization) \cite{Aiello09, Fainman84, Kis05}. The intensity admixture $I_s$ of $s$-polarization in the reflected beam is indeed proportional to \cite{Merano09} (Supplementary information) 
\begin{equation}
I_s\propto\bigg(\frac{\theta_y}{tan\hspace{0.5 mm}\theta}\bigg)^2
\end{equation}
where $\theta_y$ is the beam aperture in the y direction (see Fig. 1). In practice the matrix describing the reflection process is not anymore $F$ and our pre-selection scheme for the input polarization is not anymore valid.
But if in place of a Gaussian beam we use a sheet Gaussian beam we avoid this problem.  

Figure 4 reports the experimental results for the AGH when we use a sheet Gaussian beam. We experimentally produce this beam by focusing the light with a cilindrical lens (focal lenght = 5 cm). In this set of measurements $w_0$= 19$\rm \mu m$, $\epsilon$ = 0.05 - 0.23 rad and z = 24 cm. Now the agreement in between theory and experiment is again excellent.
  
In conclusion, we reported the weak measurement of the AGH. Our experimental method is successful in faithfully amplifying the AGH shift at any angle of incidence. Our set up is an optical analog \cite{Sudarshan89, Ritchie91} of the original proposal of Aharonov, Albert and Vaidman \cite{Aharonov88} as soos as we recognise the reflection matrix as part of the pre-selection process. Because of Brewster cross polarization close to $\theta_B$ the method is still valid as long as we use a Gaussian sheet beam instead of a Gaussian beam. This last choice is not a conceptual limitation in the spirit of the original weak measurement proposal, where a fully scalar Gaussian beam was used.
  
\begin{figure}
\includegraphics{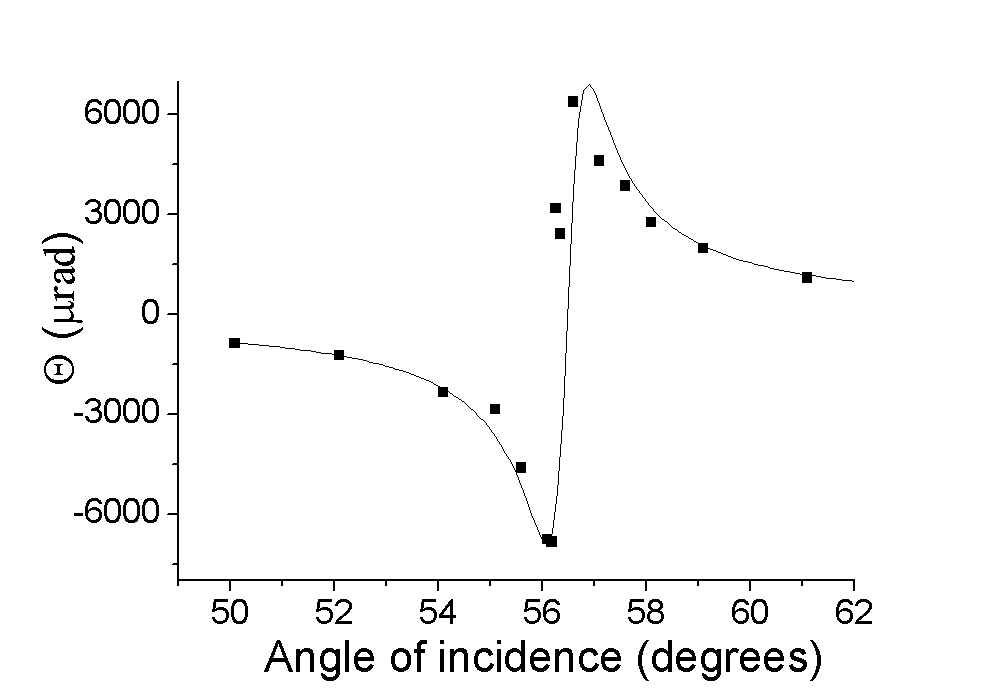}
\caption{Angular Goos-H$\rm\ddot{a}$nchen shift for a $p$ polarized sheet Gaussian beam with respect to a $s$ polarized one. Solid dots are experimental data, the line is the theoretical prediction. The agreement in between theory and experiment is excellent.}
\end{figure}

\begin{acknowledgements}
We thank to Giorgio Delfitto for his technical assistance.We gratefully acknowledge the support from PRAT-UNIPD contract (CPDA),118428 and Fondazione Cassa di Risparmio di Padova e Rovigo (CARIPARO) for their financial aid. 
\end{acknowledgements}

%\bibliography{letter}
%\bibliographystyle{osajnl}

\end{document}